\documentclass{PoS}
\usepackage{amsmath}

\title{QED-corrected Lellouch-L{\"u}scher formula for $K \rightarrow \pi\pi$ decay}

\ShortTitle{QED-corrected Lellouch-L{\"u}scher formula for $K \rightarrow \pi\pi$ decay}

\author{\speaker{Yiming Cai}
\\
        Maryland Center for Fundamental Physics and Department of Physics,\\
University of Maryland, College Park, MD 20742, USA\\
        E-mail: \email{yiming@umd.edu}}

\author{Zohreh Davoudi\\
         Maryland Center for Fundamental Physics and Department of Physics,\\
University of Maryland, College Park, MD 20742, USA, and\\
RIKEN Center for Accelerator-based Sciences,
Wako 351-0198, Japan\\
        E-mail: \email{davoudi@umd.edu}}

\abstract{A precise SM prediction for the direct CP violation in the $K \rightarrow \pi\pi$ decay process is of great importance in confronting experiments and constraining new physics. The state-of-art lattice QCD study of this process will soon achieve a precision that QED effects can no longer be neglected. The inclusion of QED in such calculations is planned, and the formalism to relate the finite-volume matrix element obtained from these calculations to the physical amplitude is underway. Here, we report on the progress towards an extension of the Lellouch-L{\"u}scher formalism in presence of QED, with the goal of enabling the extraction of physical amplitudes for the $K\rightarrow \pi\pi$ process with charged initial and/or final states.}

\FullConference{The 36th Annual International Symposium on Lattice Field Theory - LATTICE2018\\
		22-28 July, 2018\\
		Michigan State University, East Lansing, Michigan, USA.}

\begin{document}

\section{Introduction}

CP violation is of great importance in understanding fundamental questions about the universe. While such a breaking appears in the weak sector of the Standard Model (SM), the amount is too small to account for the observed matter-antimatter asymmetry in the matter-dominant universe. Thus, new mechanisms beyond the SM is required to explain this imbalance. To test the predictions from various new-physics models, a precise determination of CP violation within the SM would be the desired baseline. In experiment, meson systems (K, B and D mesons) are often used to detect CP-violating phenomena. Because of its smaller mass, kaon is created in high-energy colliders more often, and once decayed, produces a cleaner final-state signal, consisting of a few pions. In kaon system, the ratio between direct CP violation parameter $\epsilon'$ and indirect CP violation parameter $\epsilon$ is quite sensitive to new physics because of its small size, which experimentally is found to have the value $\text{Re}(\epsilon'/\epsilon)= 16.6(2.3) \times 10^{-4}$~\cite{Batley:2002gn, Abouzaid:2010ny, Patrignani:2016xqp}. Over the past few years, theoretical determinations of this ratio have become possible with the advancements in first-principles lattice quantum chromodynamics (LQCD) technology. The current theoretical value obtained by the RBC and UK collaboration is $\text{Re}(\epsilon'/\epsilon)= 1.38(5.15)(4.59) \times 10^{-4}$~\cite{Bai:2015nea}. This value is produced in the limit where quarks are electrically neutral and the up and down quark masses are equal. The large uncertainties in this quantity are expected to be reduced to $\sim10\%$ within a few years. At that stage, isospin breaking effects arising from the mass difference between the light quarks and quantum electrodynamics (QED) effects may need to be taken into account.

Through a naive estimation, QED corrections should be $\mathcal{O}(\alpha) \sim 1\%$, where $\alpha$ is the QED fine structure constant. The actual size of QED corrections may, however, be larger. The ratio $\text{Re}(\epsilon'/\epsilon)$ is related to the amplitudes for decay of kaon to two pions with isospin zero, $A_0$, and that of the decay of a kaon to two pions with isospin 2, $A_2$, through: $\frac{\epsilon'}{\epsilon} = 
  \frac{i w e^{i(\delta_2 - \delta_0)}}{\sqrt{2}\epsilon}\big[\frac{\text{Im}(A_2)}{\text{Re}(A_2)}-\frac{\text{Im}(A_0)}{\text{Re}(A_0)} \big]$, where $\delta_0$ and $\delta_2$ are the two-pion s-wave scattering phase shifts with isospin zero and two, respectively. Given the enhancement of $A_0$ over $A_2$ by a factor of $\approx 22$, it is highly plausible that the QED effects, accompanied by strong interaction effects, are enhanced in $A_0$ compared with $A_2$. This means that the size of QED corrections in this amplitude could be as large as $\sim10\%-20\%$~\cite{Christ:2017pze}, comparable to the target uncertainty of the upcoming LQCD calculations. Although implementing physical masses of up and down quark is computationally challenging, such calculations are not conceptually different than those currently performed. The situation is different when it comes to the inclusion of QED in LQCD calculations, where the infinite-range of QED interactions impose difficulty in a finite volume with periodic boundary conditions (PBCs)~\cite{Duncan:1996xy, Blum:2007cy, Hayakawa:2008an, Borsanyi:2014jba, Davoudi:2014qua, Lucini:2015hfa, Endres:2015gda}. Assuming that the first LQCD+LQED calculations of $K \to \pi \pi$ can be accomplished in near future with similar improved uncertainty, a proper treatment must be necessarily adopted such that the physical infinite-volume amplitude can be obtained from the finite-volume (FV) Euclidean result. In this talk, we present a strategy to derive a QED-corrected Lellouch-L{\"u}scher formula to address this problem. For an alternative strategy, see Ref.~\cite{Christ:2017pze, Feng:2018qpx}.

\section{Background}
\noindent
The generalized Lellouch-L{\"u}scher formula for $K \to \pi \pi$ with QED combines the FV formalism for extracting $2 \to 2$ scattering amplitudes in absence and presence of QED, and $1 \to 2$ transition amplitudes in absence of QED, and extends them to $1 \to 2$ processes in presence of QED. Here, we briefly review these formalisms before introducing our strategy for obtaining the generalized Lellouch-L{\"u}scher formula.


\noindent
\textbf{$2 \to 2$ processes in absence of QED:} Eigenenergies of two-hadron states in a finite cubic volume with PBCs can be obtained from Euclidean correlation functions of two hadrons using LQCD. However, accessing the infinite-volume scattering amplitudes is a nontrivial task as scattering amplitudes cannot be obtained from infinite-volume Euclidean correlation functions~\cite{Maiani:1990ca}. To overcome this obstacle, L{\"u}scher provided a mapping between the FV spectra of two hadrons and the elastic scattering amplitudes~\cite{Luscher:1990ux}. Assuming that scattering in higher partial waves is suppressed at low-energies, L{\"u}scher's formula can be written as $\tan\delta(q^*) = \tan\phi(q^*)$, where $q^*$ is the center-of-mass momentum of two hadrons, $\phi$ is a kinematic function that depends on the spatial extent of the cubic volume, $L$, and $\delta$ is the s-wave elastic scattering phase shift. Volume corrections that scale as $e^{-RL}$ are neglected, where $R$ denotes the finite range of interactions.


\noindent
\textbf{Inclusion of QED in the single-hadron sector:}
Given the infinite range of QED interactions, there is an IR divergence that needs to be dealt with in a finite volume with PBCs. The IR divergence makes the definition of the photon propagator ill defined and violates Gauss's law. One approach to remedy this issue is to remove the spatial photon zero mode of the photon from LQCD+LQED calculations~\cite{Hayakawa:2008an, Borsanyi:2014jba, Davoudi:2014qua, Davoudi:2018qpl}, with which one can obtain the leading power-law FV corrections to the mass of hadrons: $m_V- m=\alpha Q^2\left(\frac{c_1}{2 L}+\frac{c_1}{mL^2}\right)+ \mathcal{O}\left(\frac{1}{L^3}\right)$, where $Q$ is the charge of hadron in units of $e$, $m$ is its mass, and $c_1 = -2.83729\cdots$. Alternative approaches exist with given advantages and disadvantages, including introducing a nonzero mass for the photon~\cite{Endres:2015gda}, or performing calculations using the so-called $C^*$ BCs~\cite{Lucini:2015hfa, Hansen:2018zre}. In what follows, we focus on the approach based on the removal of the zero mode of the photon, and keep FV QED corrections up to and including $\mathcal{O}(\alpha/L^2)$, so that only the universal modifications to the mass of hadrons enter the formalism. 


\noindent
\textbf{$2 \to 2$ processes in presence of QED:}
Savage and Beane provide a formalism to obtain the low-energy scattering of two charged hadrons in the nonrelativistic limit from the FV spectra~\cite{Beane:2014qha}. Aside from the removal of the spatial zero mode of the photon to restore Gauss's law, they introduce a further constraint to avoid the need for inclusion of an infinite ladder of photon exchanges in the coulomb scattering of two charged particles. Given an IR enhancement, these contributions need to be taken into account to all order at low energies. However, since the lowest momentum mode of the photon in the regulated theory is $\frac{2 \pi}{L}$, if the spatial extent of the volume obeys $L\ll \frac{2\pi}{\alpha m}$ ($m$ is the mass of identical hadrons), one remains in the perturbative regime, i.e., it suffices to include only $\mathcal{O}(\alpha)$ contributions, those arising from both potential and radiative photons. Further, contributions in which a radiated photon is absorbed after the charged hardons have interacted multiple times through strong or QED  interactions are shown to be suppressed by powers of $\sim \sqrt{m/L}$ in the nonrelativistic case~\cite{Beane:2014qha}.

The modified L\"uscher's formula in presence of QED can be written as
\begin{align}
    -\frac{1}{a'_C} + \frac{1}{2}r'_0 p^2= \frac{1}{\pi L}S^C(\tilde{p}) + \alpha m\left[\ln\left(\frac{4\pi}{\alpha mL}\right)- \gamma_E \right]+ \mathcal{O}\left(\alpha^2\right) ,
\end{align}
where $a'_C$ and $r'_0$ are (Coulomb-corrected) scattering length $a_C$ and effective range parameter $r$ of s-wave interactions, defined with a modified kinematic to take into account the shifted mass of hadrons as a result of QED corrections in a finite volume. Further, $S^C(x)\equiv  S(x) -\frac{\alpha mL}{4\pi^3} S_2(x) + \frac{\alpha ma^2_Cr_0}{\pi^2L^2}\mathcal{I}\big[ S(x)\big]^2 + ...$, where $S(x)$ is the L\"uscher's $Z$-functions known from $2 \to 2$ scattering formula, $\mathcal{I}=-8.913632\cdots$, and $S_2(x)$ is a new FV sum, see Ref.~\cite{Beane:2014qha}. This formula can be derived as an intermediate product of the formalism for $1 \to 2$ processes in presence of QED, given our general approach outlined below. Before moving on, one must note that there is a significant difference between L\"uscher's formula and modified L\"uscher's formula in presence of QED. The latter is not a universal relation, i.e., even though the photon zero mode is removed, the modified interaction range still does not satisfy $R < L/2$, which is a requirement in the derivation of a universal L\"uscher's formula. This nonuniversality presents itself at least in two places: i) there are QED FV corrections to the mass of scattered hadrons at $\mathcal{O}(1/L^3)$ and beyond that depend on the structure of hadrons, ii) the radiative photons (producing t and u channel loops) are sensitive to the detail of strong interactions, and the quantification of FV effects requires adopting an effective field theory description of strong interactions.


\noindent
\textbf{$1 \to 2$ processes in absence of QED:} The Lellouch-L{\"u}scher formula gives access to transition amplitudes of single-hadron decay to two hadrons in infinite volume from a corresponding FV matrix element (ME)~\cite{Lellouch:2000pv}. Since this formula is another biproduct of our derivation of the modified formula in presence of QED, as sketched below, we only remind the reader of the result for the $K \to \pi \pi$ process
\begin{align}
\frac{\left|_{\infty}\big\langle \pi(E_n/2,\vec{q}_n),\pi(E_n/2,-\vec{q}_n)| H_w |K(m_K,\vec{0})\big \rangle_{\infty}\right|^2}{\left|_{V}\big\langle \pi(E_n/2,\vec{q}_n),\pi(E_n/2,-\vec{q}_n)| H_w |K(m_K,\vec{0})\big \rangle_{V}\right|^2}
=\frac{32\pi^2L^3}{ E_n|\vec{q}_n|}(\delta'+\phi'),
\label{eq:LL}
\end{align}
presented in a frame where kaon is at rest. Here, $E_n$ is a FV eigenenergy of the final-state pions such that $m_K=E_n$. This can be achieved by tuning the volume or implementing certain BCs. $q_n$ is the corresponding on-shell momentum of each pion. Prime denotes derivative with respect to the energy. For conventions regarding the normalization of the states and the interaction Hamiltonian $H_w$, see Ref.~\cite{Christ:2015pwa}.

\section{Modified finite-volume formalism for $1 \to 2$ processed in presence of QED}
\noindent
The derivation of the QED-corrected Lellouch-L\"uscher formula will be sketched in this section, presenting only qualitative features and leaving the technical details to an upcoming publication. Our derivation follows that of Refs.~\cite{Christ:2015pwa, Briceno:2015csa} for the case of Lellouch-L{\"u}scher formula in absence of QED. As the isospin is no longer a symmetry of systems in presence of QED, we consider amplitudes for both $K^+ \rightarrow \pi^+ \pi^0$ and $K^0 \rightarrow \pi^+ \pi^-$, separately. These two cases can be obtained from various limits of a general formalism that treats both the initial one-particle and final two-particle states as electrically charged hadrons.

\noindent
\textbf{$K \to K$ correlation function and the Lellouch-L\"uscher formula:} The starting point of the derivation is to consider the correlation function for $K \to K$ in presence of both QED and QCD interactions below the inelastic threshold for producing more than two pions. Additionally, the photon energy is gapped as a result of the removal of its spatial zero mode in the finite volume. Therefore, the emittance or exchange of on-shell photons in intermediate or final states can be prevented with the requirement: $m_K - 2m_{\pi} < \frac{2\pi}{L}$. For physical masses, this constraint leads to $L\lesssim 5.5~\text{fm}$, for which $m_{\pi} L \lesssim 4$ is still sufficiently large for exponential corrections from the finite range of strong interactions to be suppressed. One last requirement, which is easily satisfied with current lattice sizes, is $L\ll \frac{2\pi}{\alpha m_{\pi}}$. As stated above, this allows for a perturbative treatment of ladder diagrams consisting of exchanged photons. The correlation function in the finite volume can be written as
\begin{eqnarray}
C^{V}_{K \to K}(t) &=& \frac{1}{2} \int \frac{dP_0}{2\pi}e^{i P_0t} \lim_{P^2 \to m_K^2} \left[\frac{P^2 - m_K^2}{Z_K}\right]^2 \lim_{T \to \infty(1-i\epsilon)} \int_{L^3} d^4x \int_{L^3} d^4y 
\nonumber\\
&& \qquad \qquad \qquad \qquad  \times ~ \langle 0 | T\left\{ K(y) \int dt_2 H_w(t_2) \int dt_1 H_w(t_1)  K^{\dagger}(x)\right\} | 0 \rangle,
\end{eqnarray}
where $Z_K$ is the overlap of the operator $K$ onto the ground-state kaon and all time integrals are taken from $-T$ to $T$. Note that one can insert energy at the weak current to allow balancing between the kaon energy and that of the two-pion FV energies, a possibility that can be easily introduced to the formalism. After inserting a complete set of states and a bit of algebra, one arrives at
\begin{eqnarray}
C^{V}_{K \to K}(t)
= L^3 e^{ i E_n t}\left|_{V}\big\langle \pi(E_n/2,\vec{q}_n),\pi(E_n/2,-\vec{q}_n)| H_w |K(m_K,\vec{0})\big \rangle_{V}\right|^2,
\label{eq:CL-I}
\end{eqnarray}
\begin{figure}[h]
    \centering
    \includegraphics[width=1\textwidth]{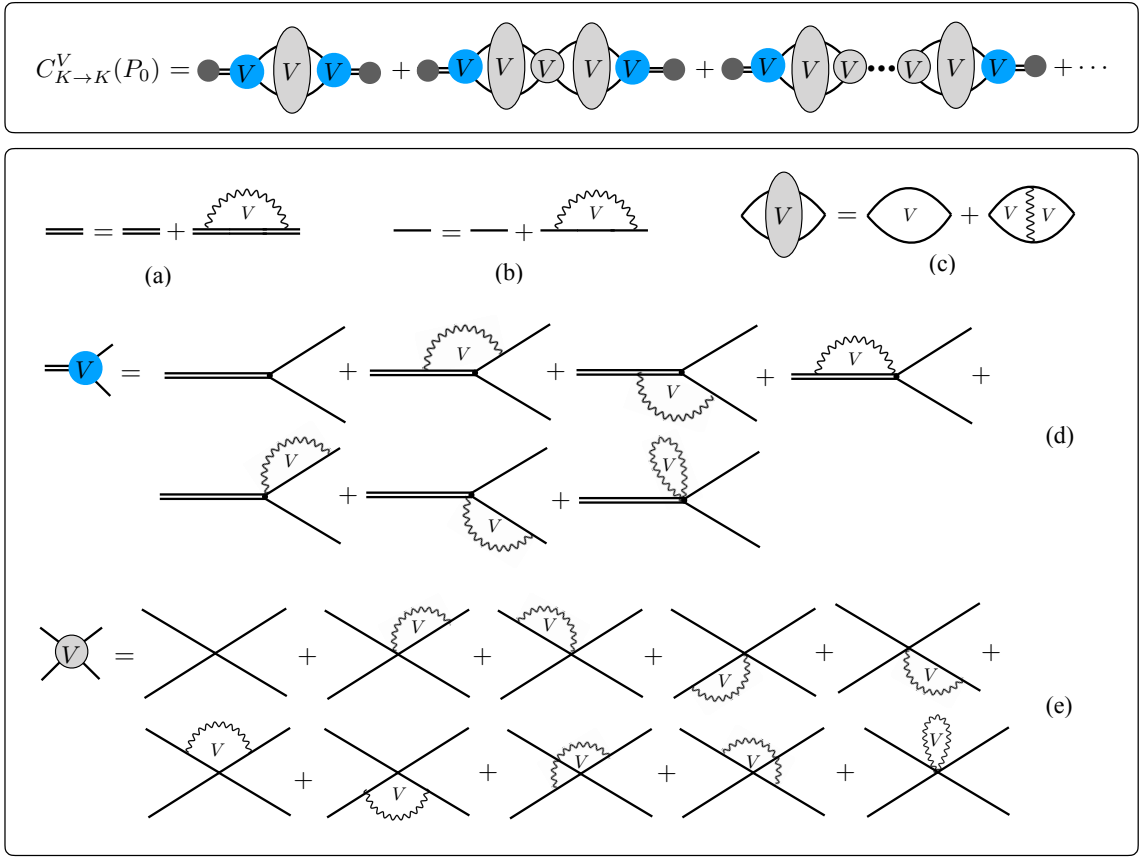}
    \caption{Dominant contributions to $C^{V}_{K \to K}(P_0)$ given the kinematic region considered in this work. $V$ inside the loops indicates that the integral over three-momenta are replaced by a discrete sum. The gray circle in the upper panel denotes the interpolating operator for the kaon. The double (single) line is the kaon (pion) propagator and the wavy line is the photon propagator. Some diagrams must be dropped depending on the channel considered. }
    \label{fig:CV}
\end{figure}
with $m_K=E_n=2\sqrt{\vec{q}_n^2+m_{\pi}^2}$ for kaon at rest. The same correlation function can be formed from the momentum-space correlation function, 
\begin{eqnarray}
C^V_{K \to K}(t) 
=\int \frac{dP_0}{2\pi}e^{i P_0t} C_L(P_0)
= e^{ i E_n t} R(E_n) \left| {_{\infty}\langle} \pi(E_n/2,\vec{q}_n),\pi(E_n/2,-\vec{q}_n)| H_w |K(m_K,\vec{0})\rangle_{\infty}\right|^2
\label{eq:CL-II}
\end{eqnarray}
where $R(E_n)$ is related to the residue of the FV correlation function for $\pi\pi \to \pi \pi$ with the inclusion of QED, as well as the QED FV modifications to $K \pi \pi$ electroweak Hamiltonian, evaluated at $E_n$. Some of these contributions, as will be commented below, are IR divergent, requiring further care in their treatment. Equating Eqs.~(\ref{eq:CL-I}) and (\ref{eq:CL-II}) obtains the modified Lellouch-L\"uscher formula. Below, we will outline the diagrammatic expansion of the momentum-space correlation function $C^{V}_{K \to K}(P_0)$ to be used in obtaining the modified Lellouch-L\"uscher formul in Eq.~(\ref{eq:CL-II}).

\noindent
\textbf{Diagrammatic expansion of $C^{V}_{K \to K}(P_0)$:} With the kinematic limits discussed above, diagrams that introduce power-law FV corrections to the correlation function, up to and including $\mathcal{O}(\alpha/L^2)$, are shown in Fig.~\ref{fig:CV}. The upper panel of the figure is a an infinite sum of s-channel $2 \to 2$ processes, with modified $K \to \pi\pi$ and $\pi\pi \to \pi\pi$ vertices due to QED interactions, as plotted in the lower panel. All propagators must be understood as QED-modified propagators, and the modified s-channel loop includes the photon exchange diagram, see Fig.~\ref{fig:CV}-a,b,c. For consistency, $C^{V}_{K \to K}(P_0)$ must be expanded to $\mathcal{O}(\alpha)$ in the end. $K\pi\pi$ and $\pi\pi\pi\pi$ vertices can be specified using chiral perturbation theory, but for the sake of generality, they can be kept arbitrary smooth functions of momenta. For completeness,  diagrams corresponding to charged initial, final and intermediate states are depicted and certain diagrams should be discarded when $K^+ \to \pi^+\pi^0$ or $K^0 \to \pi^+\pi^-$ process is considered. All these diagrams are evaluated in a finite cubic volume with PBCs and with the zero mode of the photon removed, using the techniques developed in Refs.~\cite{HASENFRATZ1990241, Davoudi:2018qpl}. The results will be presented in an upcoming publication.

\begin{figure}[h]
    \centering
    \includegraphics[width=0.8\textwidth]{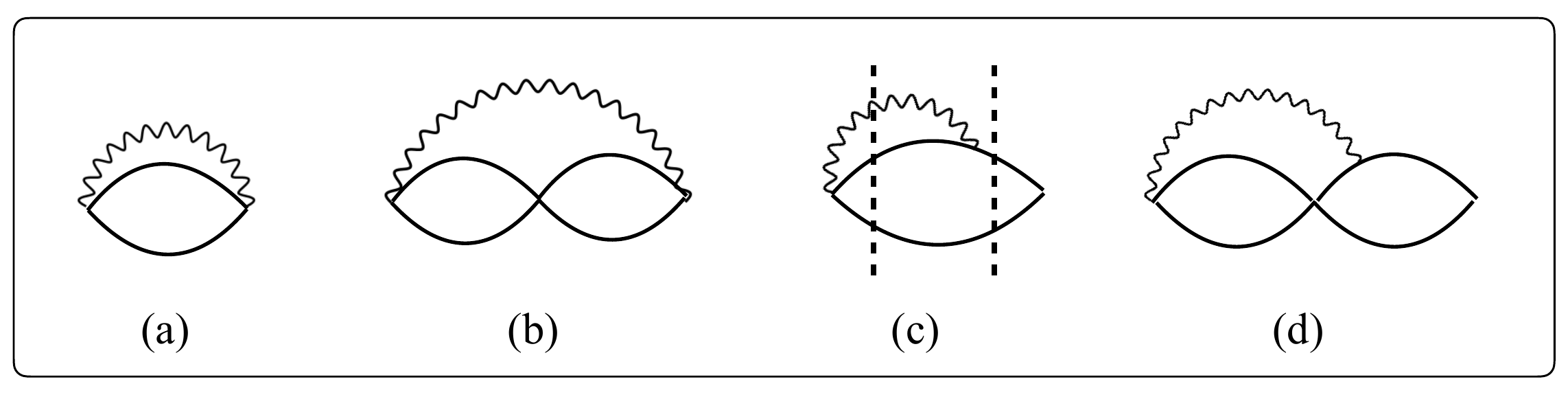}
    \caption{QED FV corrections to the diagrams in (a), (b) and (d) are of higher order in the large-volume expansion than that is considered in this work. Diagram in (c) has already been included in the expansion of $C^V_{K \to K}(P_0)$ in Fig.~\ref{fig:CV}.}
    \label{fig:others}
\end{figure}

\noindent
\textbf{Suppressed contributions in a finite volume at $\mathcal{O}(\alpha)$:} Diagrams shown in Fig.~\ref{fig:CV} are not the only contributions at $\mathcal{O}(\alpha)$. The remaining contributions at this order are those that involve a photon traveling over one or more than one s-channel two-pion loops. Examples of such contributions are shown in Fig.~\ref{fig:others}-a,b,d. Given the kinematic limit imposed, none of the particles running in the loops can go on shell. It is then straightforward to show that the QED FV corrections to these diagrams are higher than $\mathcal{O}(\alpha/L^2)$, which we are not keeping track of. As a result, contributions such as Fig.~\ref{fig:others}-a,b,d or those with a photon traveling over any number of s-channel two-pion loops can be neglected. Finally, let us remind that the volume corrections due to strong interactions are exponentially suppressed unless the intermediate particles in loops can go on shell. This implies that the effect of diagrams such as that in Fig.~\ref{fig:others}-c are already captured by the modified $K \pi\pi$ or $\pi\pi\pi\pi$ vertices in Fig.~\ref{fig:CV} coupled to s-channel two-pion loops, where both the vertex and the loop are evaluated at on-shell kinematics.

\noindent
\textbf{On the infrared divergence:} Diagrams that involve a photon connecting initial and final states can be seen to introduce an infrared divergence. Such divergence is absent from the full FV contributions as the photon zero mode is already regulated. The divergence, however, shows up when one tries to separate the infinite-volume and the purely FV parts of the sums, see e.g., Ref.~\cite{HASENFRATZ1990241}. It is well known that the physical amplitude for $K \to \pi\pi$ suffers from this IR divergence, and that taking into account the $K \to \pi\pi\gamma$ process with the final-state photon below the detector resolution $\Delta E$ cures this problem~\cite{Cirigliano:1999ie, Cirigliano:2003gt}, see also Ref.~\cite{Lubicz:2016xro, Tantalo:2016vxk, Giusti:2018guw} for IR divergence in the context of a FV formalism for leptonic decays. An extra step of the derivation of the FV formalism is to define a scheme in which the cross section for $K \to \pi\pi$, added to that of $K \to \pi\pi\gamma$, can be matched to the FV ME for $K \to \pi\pi$. Note that with our kinematic choice, the photon cannot be produced as an on-shell state, resembling a scenario in which the detector resolution for detecting a photon in infinite volume is $\Delta E = 2\pi/L$. The proper matching of the FV MEs and physical IR-finite cross sections is currently under development and will be presented in an upcoming publication.
  
\section{Conclusion and outlook}
\noindent
This talk reported on the progress towards building a mapping between the FV MEs obtained from LQCD+LQED calculation and the physical MEs for the $K \to \pi\pi$ process, a necessary step towards improving the precision of the SM determination of the important CP-violation parameter $\text{Re}(\epsilon'/\epsilon)$. This mapping is far more complicated compared with the regular Lellouch-L{\"u}scher formula for $1 \to 2$ processes, given the modifications from photon exchanges and radiations at the level of the strong and weak interaction vertices, as well as an inherent IR divergence in the physical amplitude. Our strategy, along with a schematic derivation of this formalism were presented in this talk, and much of the technical detail and the final results are left to an upcoming publication.

\begin{acknowledgments}
\noindent
Y.C. acknowledges the support of the U.S. Department of Energy. Z.D. was partly supported by the Maryland Center for Fundamental Physics.
  \end{acknowledgments}

\bibliographystyle{JHEP}
\bibliography{bibi}

\end{document}